\documentclass[]{aa}
\usepackage{txfonts,graphicx,amssymb,subfigure}
\usepackage[normalem]{ulem}
\usepackage{xcolor}

\begin{document}
\title{Galactic winds and the origin of  large-scale magnetic fields}

   \author{D.~Moss\inst{1}
          \and
        D.~Sokoloff \inst{2} }

   \offprints{D.~Moss, david.moss@manchester.ac.uk}

   \institute{ School of Mathematics, University of Manchester, Oxford Road, Manchester, M13 9PL, UK
   \and
    Department of Physics, Moscow University, 119992 Moscow, Russia}

   \date{Received ..... ; accepted .....}

\abstract{Observations of dwarf galaxies suggest the presence of large-scale magnetic fields.  However the size and slow rotation of these galaxies appear
insufficient to support a mean-field dynamo action to excite such fields.
}
{Here we suggest a new mechanism 
to explain large-scale magnetic fields in galaxies that are too small to support mean-field dynamo action. 
The key idea is that we do not identify large-scale and mean magnetic fields. 
In our scenario 
the the magnetic structures originate from a small-scale dynamo which produces small-scale magnetic field in the galactic disc and a
galactic wind that transports this field into the galactic halo where the large turbulent diffusion increases the scale and order of the field.
As a result, the magnetic field becomes large-scale; however its mean value remains vanishing in a strict 
sense.
}
{We verify the idea by numerical modelling of two distinct simplified configurations, a thin disc model using the no-$z$ approximation, and an axisymmetric model using cylindrical $r,z$ coordinates.}
{Each of these allows reduction of the problem to two spatial dimensions.
Taken together, the models support the proposition that the general trends will
persist in a fully 3D model.
We demonstrate that a pronounced large-scale pattern can develop in the galactic halo for a wide choice of the dynamo governing parameters.
}
{We believe that our mechanism  can be relevant to explaining the presence of the
fields observed in the halos of dwarf galaxies. We emphasize that detailed modelling of the proposed scenario needs 3D simulations,  and adjustment to the 
specific dynamo governing parameters of dwarf galaxies.}


\maketitle

\titlerunning{Winds and large-scale magnetic fields}
\authorrunning{Moss \& Sokoloff}

\section{Introduction}

A variety of celestial bodies, such as galaxies, stars and planets contain large magnetic fields (e.g. R\"udiger et al. 2013). The conventional mechanism
believed to be responsible for their origin is a dynamo, 
driven by differential rotation 
and the  mirror-asymmetric motions of an electrically conducting medium. 
This kind of dynamo action results in mean 
magnetic field generation and the large-scale (regular)\footnote{Linearly polarized emission traces ``ordered'' magnetic
fields, which can be either ``regular'' (or large scale) magnetic fields
(preserving their direction over large scales) or ``turbulent anisotropic''
fields (with multiple field reversals).} magnetic field  of these objects
is identified with this mean field. Of course, many 
details of the mechanism remain debatable. However models of dynamo generation for specific celestial bodies 
are gradually becoming more realistic. 

The dynamo mechanism is a threshold phenomenon, so dynamo drivers have to be strong enough to overcome various 
magnetic field losses. The strength of dynamo action, measured by a dimensionless number (the dynamo number) is 
determined partially by the size of the body in question. Celestial bodies are much larger than laboratory 
devices and magnetic field excitation by a dynamo occurs much more easily in celestial bodies than in the laboratory.

There are however celestial bodies that are relatively small 
and  rotate slowly compared with other bodies of similar type,
and nevertheless contain large-scale magnetic fields. Dwarf galaxies which 
are substantially smaller than spiral galaxies and nevertheless demonstrate 
the presence of large-scale 
magnetic fields, e.g. by showing polarized synchrotron emission; Chy\.zy et al. (2016)
provide a straightforward example (NGC 2976). The important point here is that dynamo 
action in 
spiral galaxies is believed (e.g. Beck et al. 1996) to be slightly supercritical (i.e. the dynamo number is 
slightly larger than 
its critical value). It is natural to expect the dynamo number in dwarf galaxies to be several times smaller 
than that  in spirals, and so below the threshold value, implying that the 
conventional dynamo explanation for large-scale magnetic field  generation
becomes difficult to maintain  (e.g. Mao et al. 2008).

Of course, estimates of the strengths of  dynamo drivers in galaxies are far 
from precise and it is possible to
play with numbers in an attempt to resolve the problem. In particular, Chy\.zy et al. (2016) show how this might be achieved in 
the case of NGC 2976. In spite of the fact that the dynamo 
active region is supposed to be located within a galactocentric radius of 2 kpc (compared with about 10 kpc in spiral 
galaxies) and that the rotation velocity is taken as 71~km~s$^{-1}$ 
(against 250 km s$^{-1}$ in spirals), they 
found the intensity of dynamo action (measured by the dynamo number) 
to be about half that in  the
Milky Way, but still larger than the critical value at which dynamo 
generation occurs. An important role in 
their estimate is played by the fact that a cosmic ray driven dynamo (Hanasz et al. 2009, Gaensler
et al. 2005, Kepley et al. 2010, Mao et al. 2012) is modelled. 
Another attempt of this kind was made by 
Siejkowski et al. (2014) who emphasized that "the environment of a dwarf galaxy 
is unfavourable for the large-scale dynamo action because of the 
very slow rotation that is required to create the regular component of the 
magnetic field", but who however argue that the intensity of dynamo drivers in 
dwarfs might be still sufficient for dynamo action. 
We recognize the possibilty, however the doubts 
suggested by Shukurov (2007) concerning the efficiency of dynamo action in 
dwarfs still persist, and it looks 
desirable to have an alternative mechanism available to explain the origin of
ordered magnetic field in dwarf galaxies.

%However this approach does not look very convincing. 
The aim of this paper is to suggest such an alternative 
approach to the problem. The key idea here is that we
do not now identify the large-scale magnetic field of a body with the 
mean field obtained by averaging taken 
over an ensemble of turbulent (convective) motions. Another important point 
is that generation of mean 
magnetic field is not the only possibility for dynamo action 
in a turbulent conducting medium. The 
additional mechanism is the small-scale (or turbulent) dynamo (e.g. Zeldovich et al. 1990;  and in particular  it operates in supersonic situations, e.g. Federrath et al. 2014).
Differential rotation and/or 
mirror asymmetry are not required, 
and turbulent (or convective) motions alone are sufficient.
The small-scale dynamo mechanism produces magnetic fields with zero mean. 
The spatial scale of such fields is comparable with the 
basic scale of turbulence in the domain where  the dynamo action occurs -- 
hence the term  small-scale dynamo. This small-scale dynamo is not so dependent 
on the size of the body, and it looks plausible that it could 
be efficient in dwarf galaxies as well as in spirals. 

%______________________________________________________________________
\begin{figure*}
\begin{center}
a \includegraphics[width=0.41\textwidth]{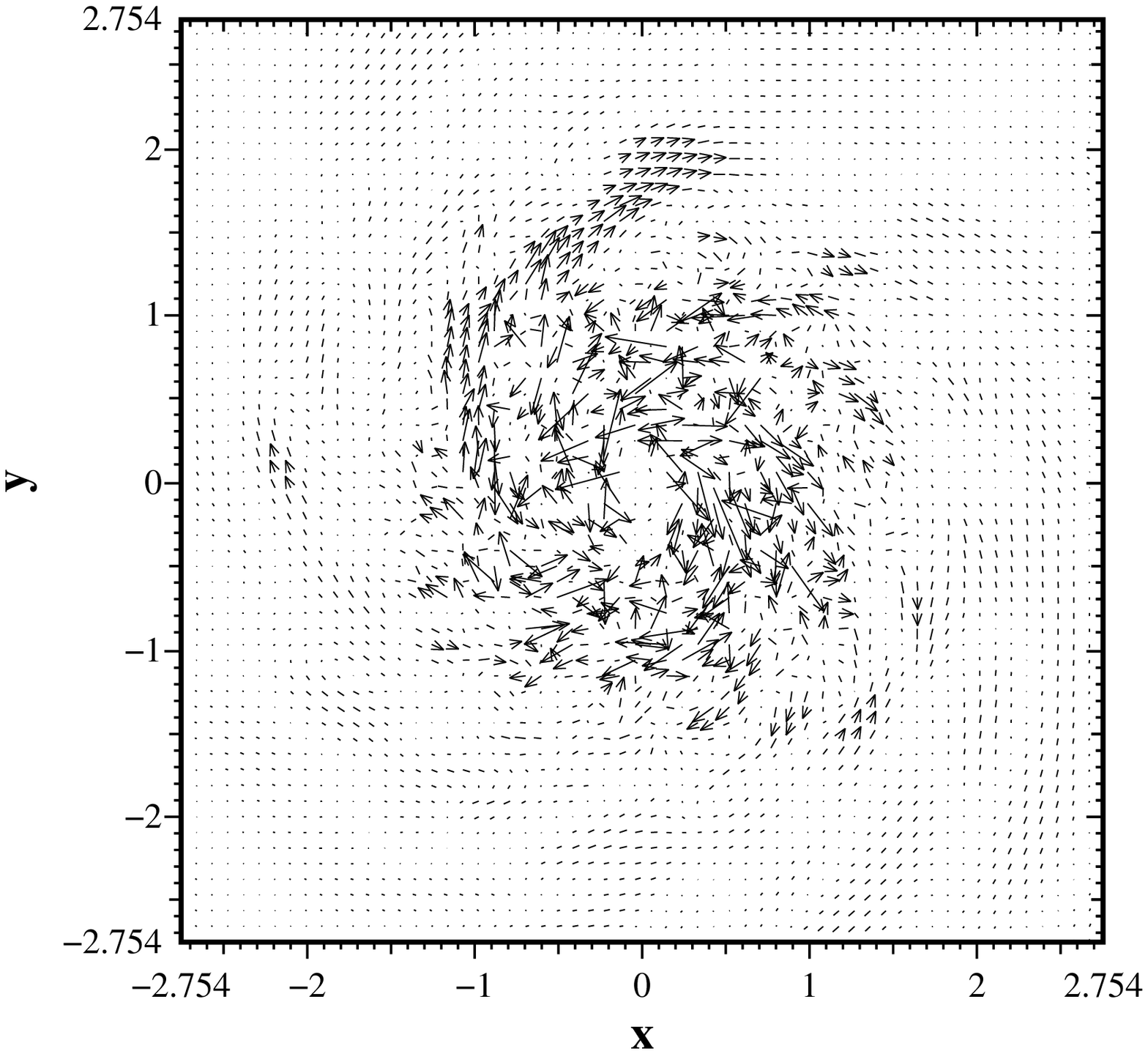}
b \includegraphics[width=0.41\textwidth]{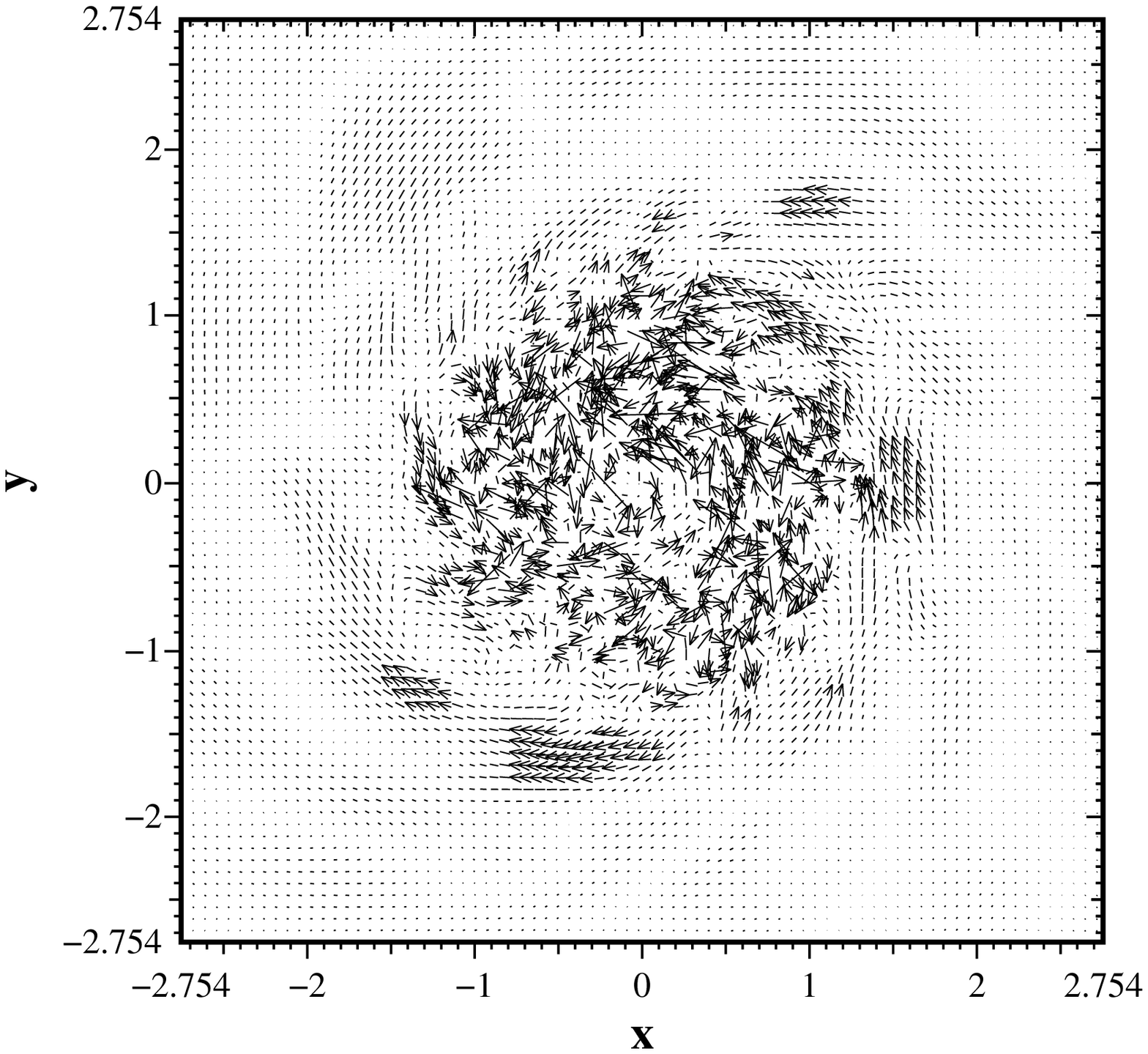}\\
c \includegraphics[width=0.41\textwidth]{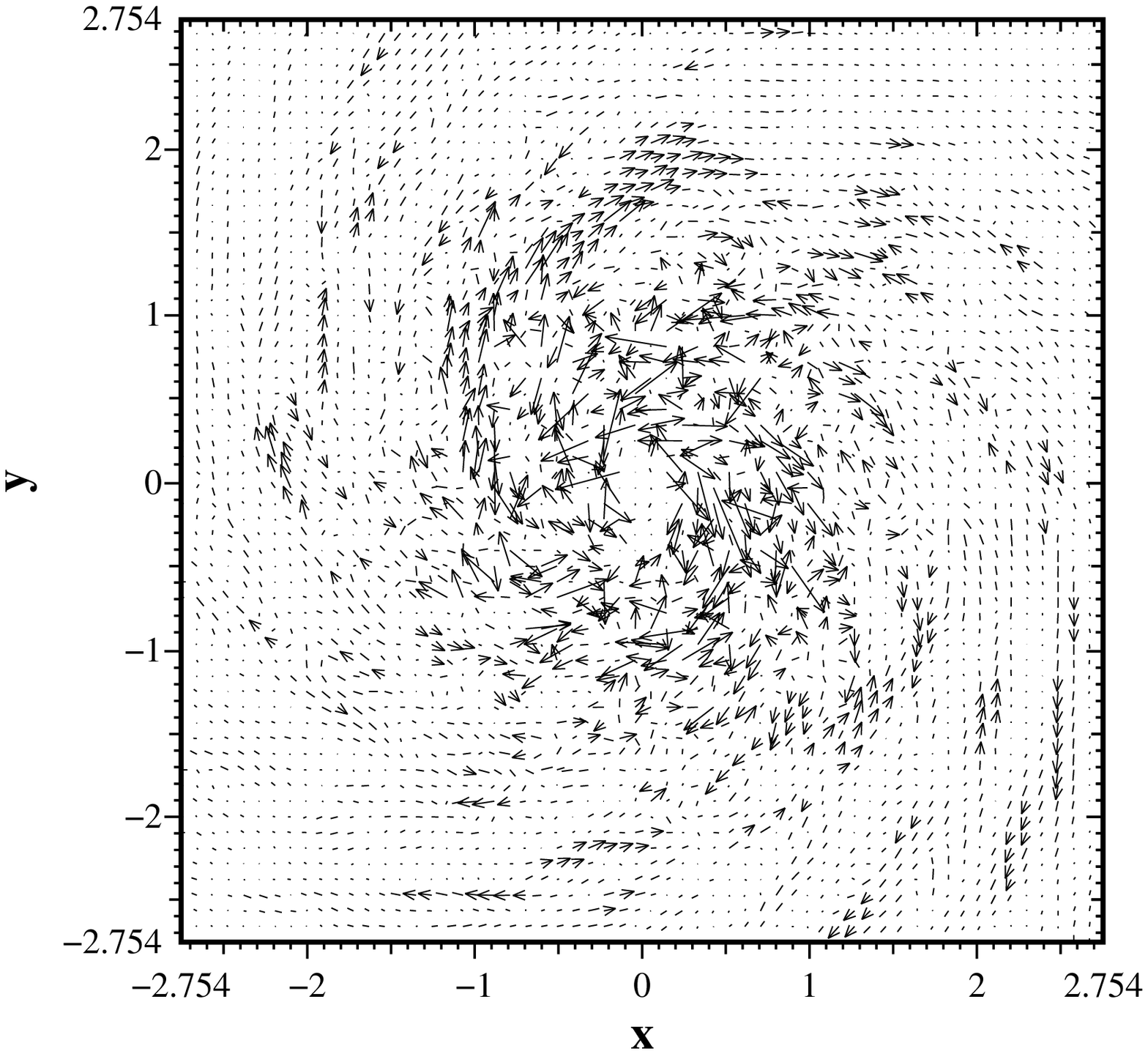}
d \includegraphics[width=0.41\textwidth]{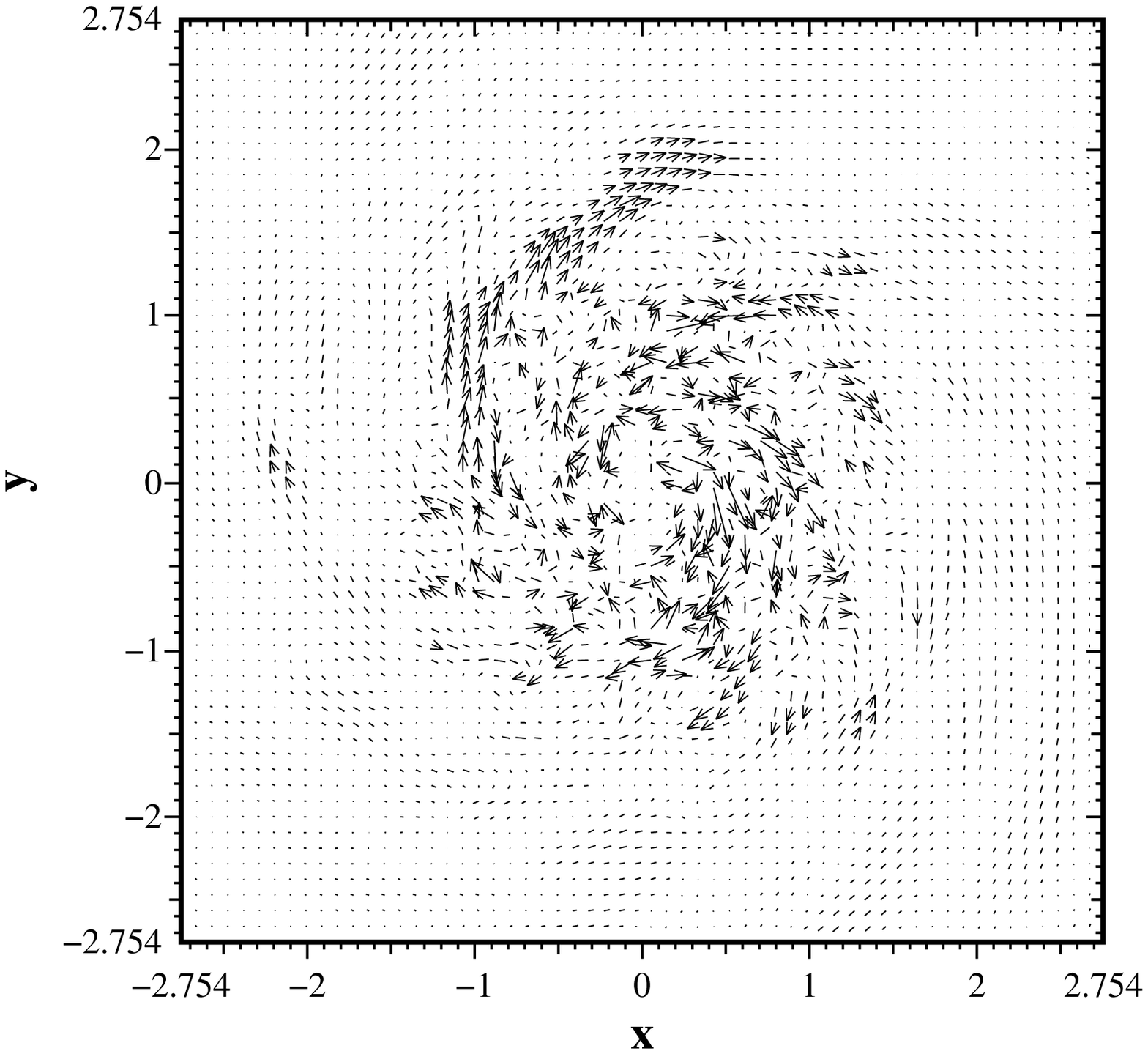}
\end{center}
\caption[]{Stationary magnetic field distributions for 
several models, all $R_\alpha=1, R_\omega=7.5, R_u=120$.
a) $\eta_h=25$, b) $\eta_h=50$, c) $\eta_h=1$; d) the model shown in panel a) with the field smoothed over a scale $0.1 R_{\rm disc}$.}
\label{fields}
\end{figure*}
%______________________________________________________________________

A further important detail of the mechanism proposed below is that a 
galactic wind can transport small-scale magnetic 
field into the media surrounding the disc of dwarf galaxies. 
The diffusivity of the  turbulent medium surrounding the disc 
is believed to be substantially larger than that in the disc (e.g. Beck et al. 1996) 
and the spatial scale of any magnetic field 
transported by the wind from the disc will become substantially larger. 
Such a field remains a field with zero mean 
in sense that it is determined by an  interplay of 
random factors.  However it will appear as a large-scale field on the
scale of galactic discs. Because we see only snapshots of magnetic field
distributions such a field would be
referred to as large-scale from the observational viewpoint. We attempt
to illustrate these ideas by computing two quite distinct dynamo models,
using first the thin disc (no-$z$) approximation (Sects.~\ref{model} 
and ~\ref{res}), and then by assuming axisymmetry (Sect.~\ref{axi}).

We emphasize that we are not attempting to model dwarf galaxies (or any other specific class of object) explicitly.
The aim of this paper is to demonstrate the idea by a simple model 
which contains the basic features 
of the proposed mechanism. Thus we have not adapted our model to the specific 
quantities (dynamo numbers, dimensions) relevant to dwarf galaxies,
leaving this to a later paper, and concentrate here on proof of concept.                   
Correspondingly, we do not attempt to give here a review of the properties of these objects.
  
%______________________________________________________________________
\begin{figure*}
\begin{center}
a \includegraphics[width=0.25\textwidth]{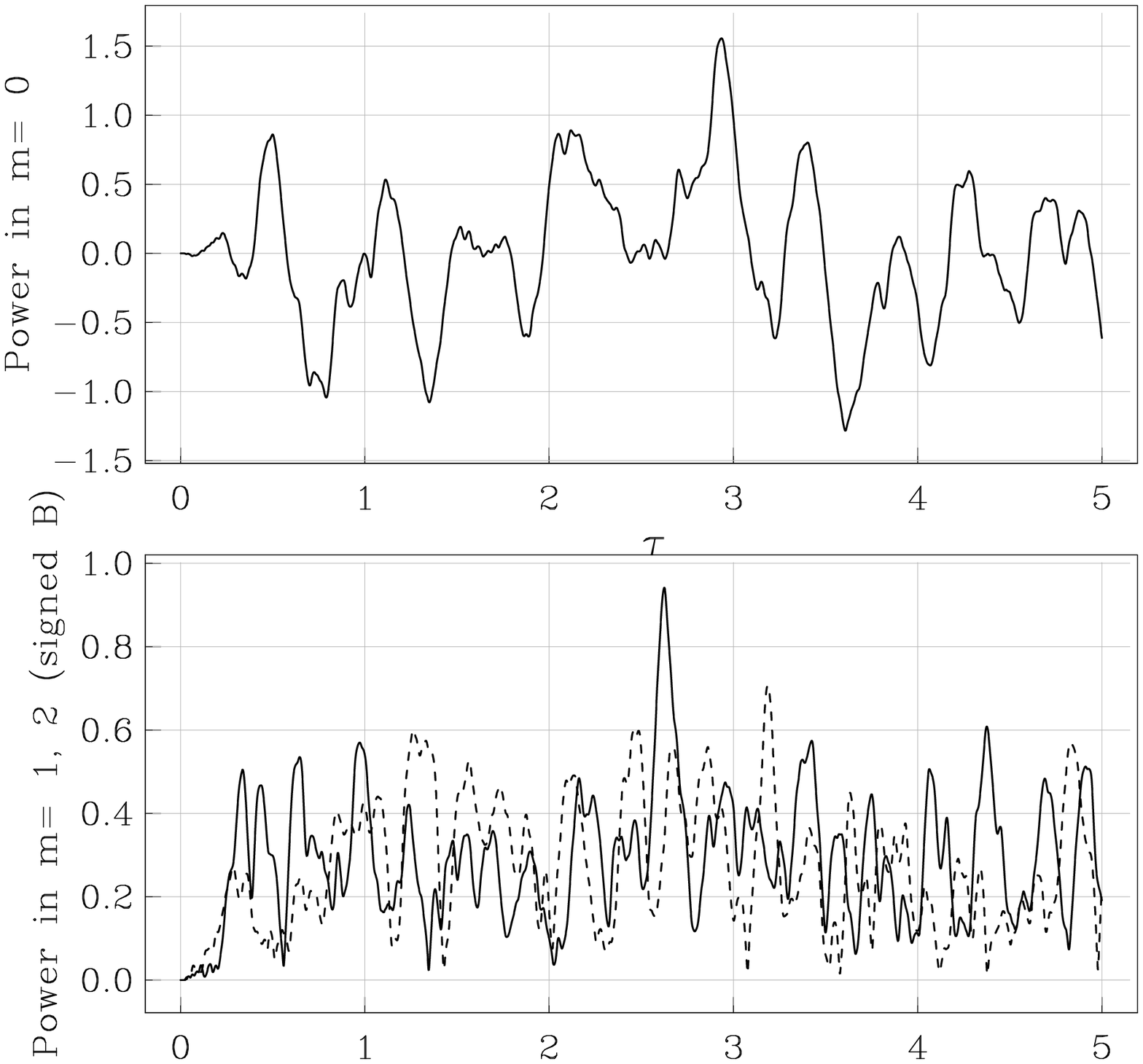}
b \includegraphics[width=0.25\textwidth]{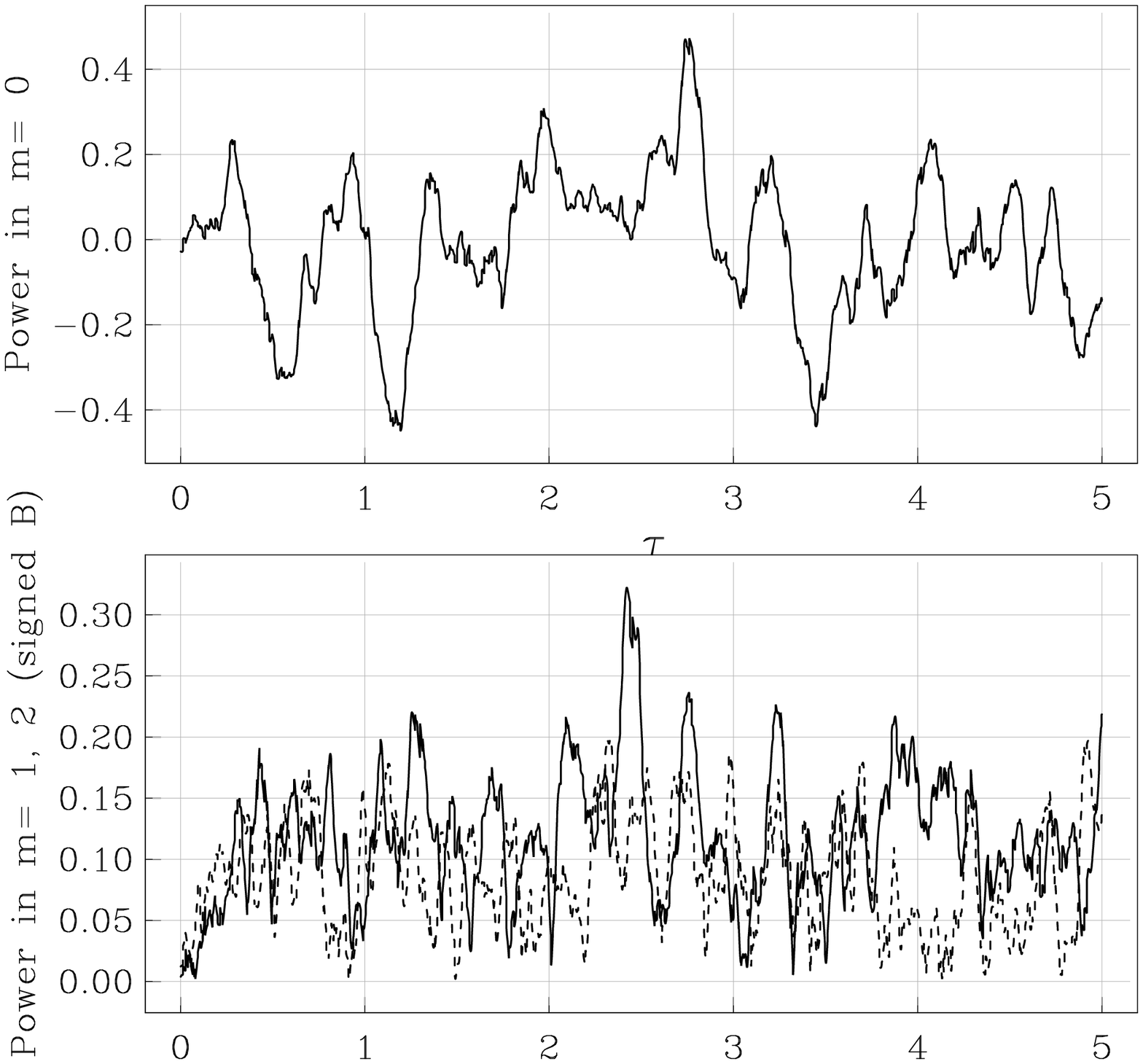}
c \includegraphics[width=0.25\textwidth]{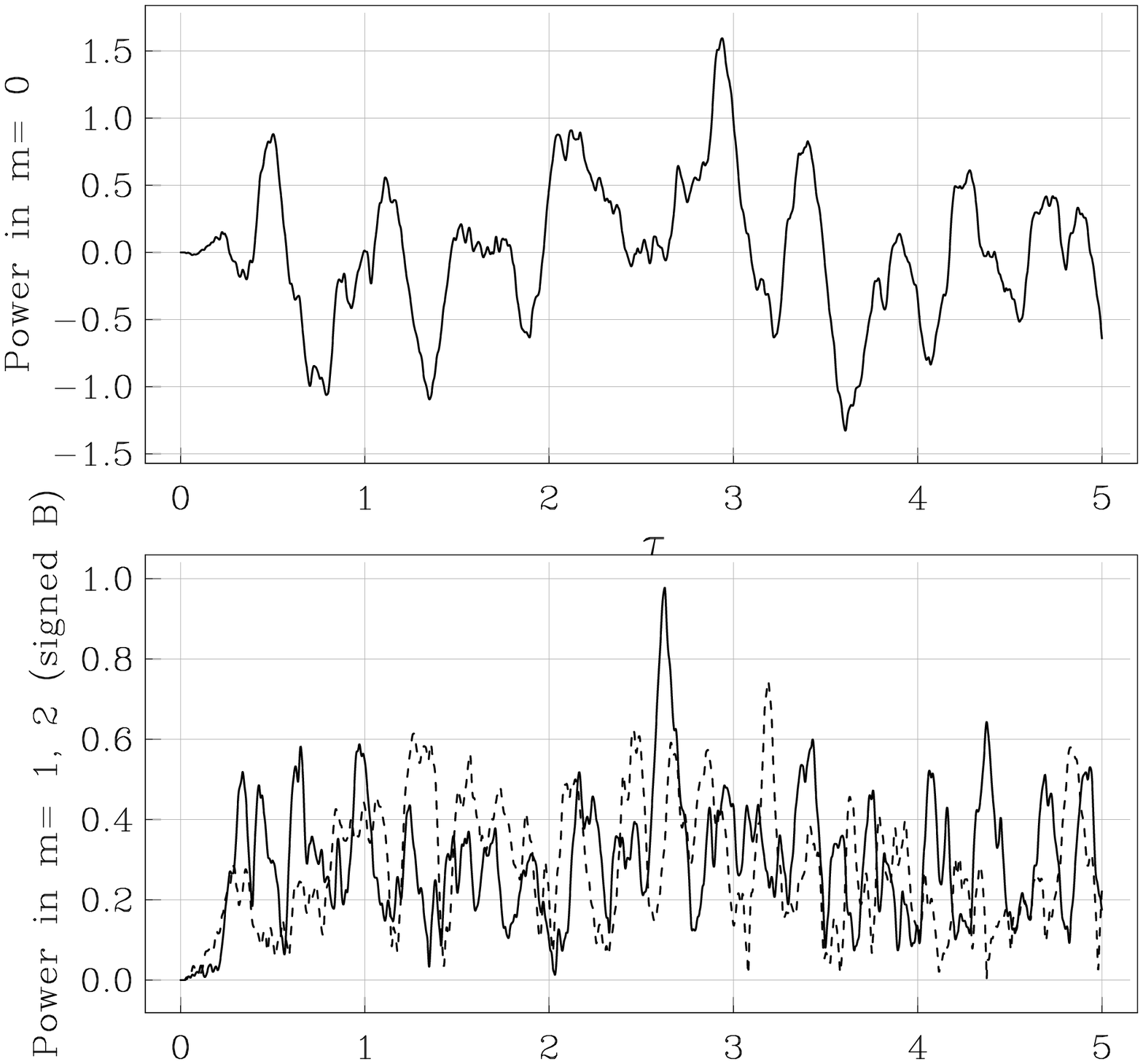}
\end{center}
\caption[]{Evolution of Fourier components $m=0,1,2$ for: a) the external field of the model
of Fig.~1a with $\eta_h=25$, b) the internal field of this model, 
c) the external field of the model shown in Fig.~1c ($\eta_h=1$).
In each case the upper panel shows the power in mode $m=0$, and the lower panel that
in modes $m=1, 2$ (solid and broken curves respectively).}
\label{temporal}
\end{figure*}

\section{Thin disc dynamo model}
\label{model}
As a basis for our modelling we consider a thin rotating disc (the centre plane of the disc is $z=0$) 
which represents the disc of a dwarf galaxy. 
The magnetic field evolution in the disc and its surrounding is described 
using the no-$z$ dynamo
model (Moss 1995), which deals with the horizontal magnetic field components $B_x$ and $B_y$ as functions of $x,y$, while the component $B_z$ perpendicular to 
the disc can be estimated from the solenoidality condition. 
The no-$z$ model has been shown to give a reasonable approximation for 
modelling of magnetic fields in disc  galaxies and, being two-dimensional, 
allows extensive modelling using reasonably limited computational facilities. 

We assume that the disc rotates differentially, and that the rotation curve is 
asymptotically  flat, 
as  is typical for spiral galaxies. An additional dynamo driver,  mirror asymmetric turbulence, is also formally present,
parametrized by the $\alpha$-effect. The turbulent diffusivity $\eta$ is assumed to be larger in the galactic 
halo than in the disc. The dynamo drivers are chosen to be slightly subcritical so that the magnetic field 
prescribed as an initial condition decays with time and is, in fact, irrelevant
to the final outcome.

The additional factor governing magnetic field evolution, the small-scale 
dynamo, is parametrized by magnetic 
field injections distributed randomly in $N_{\rm spot}$ "hotspots" 
(star forming regions)
in the disc, which sporadically supply small-scale magnetic field.
The injection time scale is taken to be of order 10 Myr.
The hotspots are relocated randomly at intervals comparable to the 
rotation period of the disc.
We verified that the injections and magnetic field losses rapidly 
result in a statistical steady state magnetic field distribution. 
This distribution appears as a random field in the disc.
We note that details of this mechanism are irrelevant here: we just need
ongoing excitation of small-scale magnetic field. Motivation and further details
are given in Moss et. al. (2012).

The simplest model of a wind is homogeneous and directed 
along the $x$-axis. 
Such a wind could 
originate from the motion of the galaxy through the intergalactic medium. 
We can model this explicitly.
Another possible origin 
is associated with supernova explosions and galactic fountains from 
strong starburst activity in the galaxy. 
In this case the wind would be expected to be approximately orthogonal to the 
disc plane. Initially we attempt to assess this by studying the effects of 
a radial wind in the disc plane. 
Thus this model of this wind is not
fully realistic: it is assumed to flow in
the disc plane and not approximately perpendicular to it. 
This unrealistic feature is necessary to
retain the  2D formulation of the problem: we do not see why changing
the wind direction would destroy the basic effect that we wish to illustrate.
However, to address this issue we also use another simplification of the
dynamo problem, assuming that the magnetic field is axially symmetric and
ignoring the azimuthal coordinate (Sect.~\ref{axi}) -- the "$r,z$ problem".
Of course, this approach has its separate limitations and inconsistencies.

Various components of the thin disc model have been
explored in several recent papers (e.g. Moss et al. 2012, 2015) and 
have been shown to give
realistic magnetic field configurations: 
here we combine all the above elements for the first time
to study the effects of a wind on the small-scale field.
The axisymmetric model in $r,z$ coordinates has a long history in
studying galactic magnetic fields.

\section{Results: no-$z$ model}
\label{res}

The strength of the dynamo drivers  is measured by the conventional dynamo
numbers $R_\alpha = \alpha_0 h/\eta_0$,
$R_\omega = \Omega_0 h^2/\eta_0$, $N_{\rm spot}$, $B_{\rm inj0}$ the rms 
amplitude of the injected field, and
$R_u = u_0 h/\eta_0$ the magnetic Reynolds number of the wind. 
Here $\alpha_0, \eta_0$ are the uniform values of $\alpha, \eta$ in the disc,
$h=0.05 R_{\rm disc}$ is the uniform disc thickness taken where $R_{\rm disc}$
is the disc radius. $\Omega_0$ is the central
angular velocity and $u_0$ is a reference velocity.
$\eta_h$ is the ratio of the external diffusivity to $\eta_0$ at large distances from the disc.

For most of our models we chose subcritical values for the mean-field dynamo, $R_\alpha=1$, $R_\omega=7.5$, $R_u=120$. We set $N_{\rm spot}=200$,
and took $B_{\rm inj0}$ to be the equipartition field strength, and 
studied the development of large-scale magnetic structures in the medium surrounding the disc. 

The first three panels of Fig.~\ref{fields} show the statistically steady 
magnetic field configurations obtained  with a radial wind with speed proportional to radius for various choices of $\eta_h$ -- see the captions. 
The fourth panel demonstrates that the effect of spatial smoothing hardly 
alters the distribution
of external field shown in Fig.~\ref{fields}a.
Of course, these figures are just snapshots taken from fluctuating distributions -- as indeed are the observed fields.

We verified that the large-scale configurations is robust to changes in numerical resolution,
and that the form of the final configuration is unaltered when the mean field dynamo is absent (i.e. $R_\alpha=0$) -- as the mean field dynamo is subcritical in the previous models its long term influence is negligible.

In Fig.~\ref{temporal} we show the evolution of the lower Fourier analysed components
of the external and internal fields for the model of Fig.~\ref{fields}a,
and the external field of the model of Fig.~\ref{fields}c -- these are typical.
Here we also show the time evolution of the global magnetic energy -- the cases with a radial wind appear very similar. The spikes correspond to the
small-scale field injections.

We  define the power in the $m=0$ part of the field by the global integral
$\int B_\phi dV$, 
and that in the azimuthal mode $m>0$ by
$\left((\int_V B_\phi \cos m\phi dV)^2 +(\int_V B_\phi \sin m\phi dV)^2\right)^{1/2}$.
Thus the power in $m=0$ can take negative values, that in higher modes is positive definite.

\begin{figure*}
\begin{center}
a \includegraphics[width=0.25\textwidth]{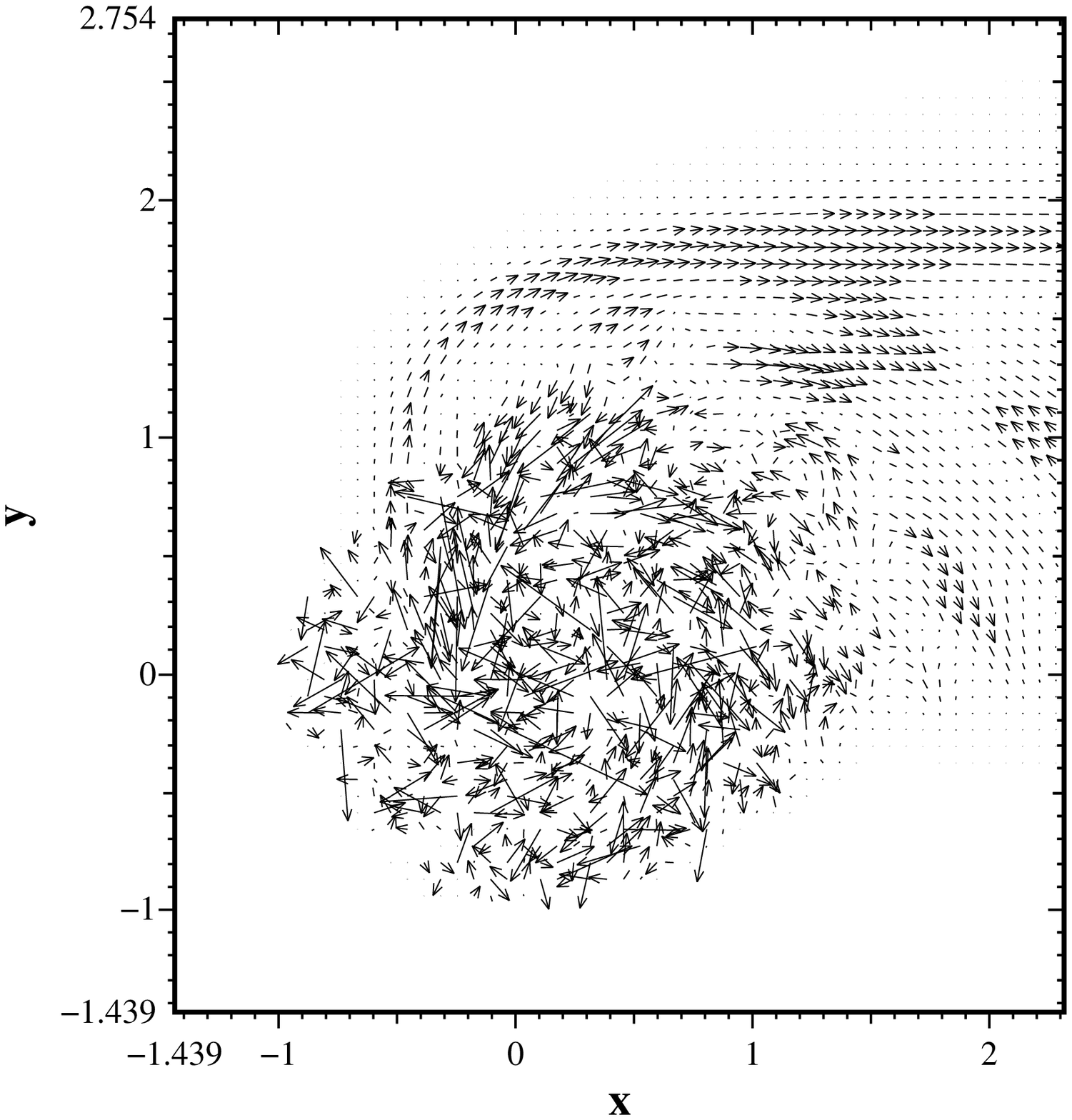}
\hspace{0.2cm}b \includegraphics[width=0.25\textwidth]{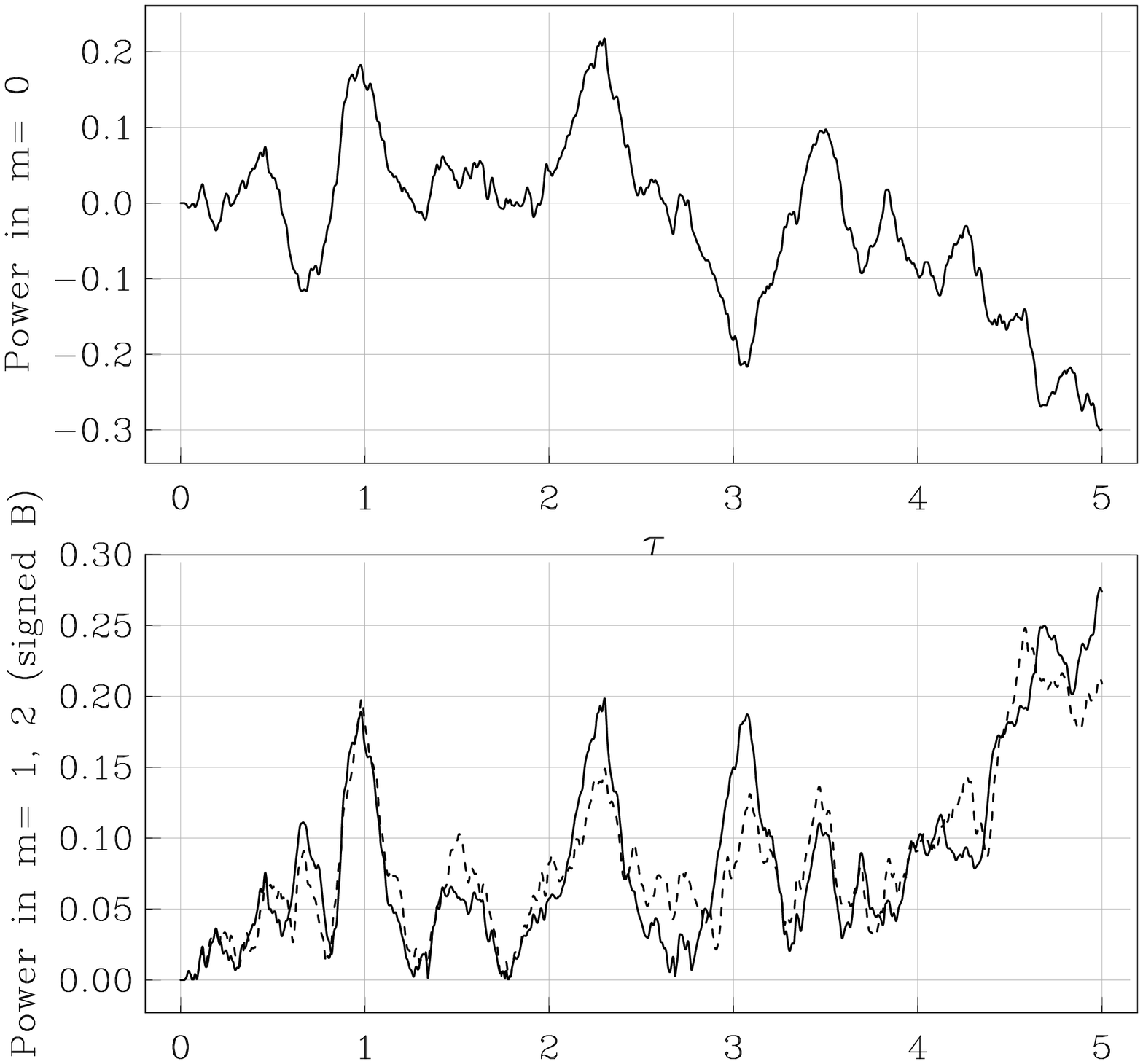}
\hspace{0.2cm}c\includegraphics[width=0.25\textwidth]{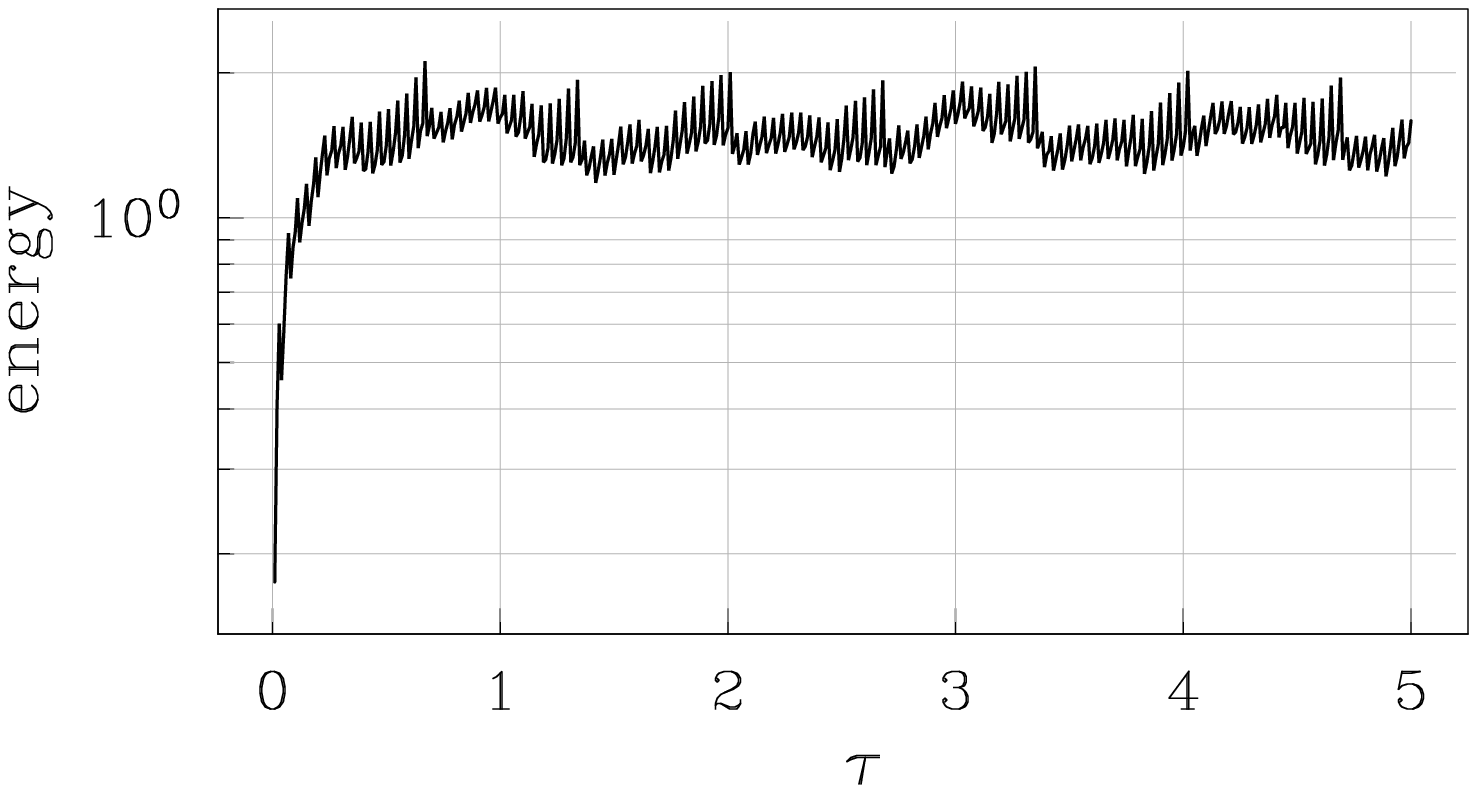}
\end{center}
\caption{The model with a uniform wind in the $x$-direction, $\eta_h=25, R_u=120$: a) the statistically steady field, b) evolution of Fourier components of the external field, c) the evolution of the global magnetic energy.}

\label{uniform}
\end{figure*}

Finally we present in Fig.~\ref{uniform} the corresponding results for a model with a uniform wind in 
the $x$-direction. Now the external field is displaced to one side of the disc.
Here we also show in panel ~\ref{uniform}c the time evolution of the 
global magnetic energy in this
case -- the examples with radial wind are very similar. The spikes
correspond to the discrete field injections

We learn from Fig.~\ref{temporal} what happens during the formation of large-scale patterns in the galactic halo. 
Indeed, we see that the mode $m=0$ becomes about 3 times larger in the external parts of the galaxy (i.e. in the 
halo) than in the galactic disc. This large-scale field fluctuates in time and is in this sense a field with zero 
mean. The large-scale component $m=0$ is present in the disc as well. However the field injections are random and have zero mean. Presumably, these effects occur because $N_{\rm spot}$ is quite, but not very large.

%______________________________________________________________________

\section{Axisymmetric model in $(r,z)$ coordinates}
\label{axi}

The experiments described in Sect.~\ref{res} using the thin disc,
no-$z$ approximation provide strong evidence
that the proposed mechanism can work in that geometry. However,
in the specific context of ordered magnetic field above the discs of dwarf 
galaxies, it might be preferable to use a more directly applicable geometry.
(Although, as we have pointed out in Sect.~\ref{res}, that geometry can be 
directly relevant to a galaxy moving through an external medium, as described in
Moss et al. 2013.) Thus we study in this section an axisymmetric galaxy model,
represented in cylindrical polar coordinates $(r, \phi, z)$ with no $\phi$-dependence. The code used is basically that of Moss et al. (1998), and
subsequently adapted for use in Moss et al. (2016). 

In these experiments we implemented the same random field injection mechanism as in the no-$z$ model of Sect.~\ref{model}, with the hotspots now being at 
randomly allocated points in $r,z$. Of course, this model also has its own
obvious deficiencies, notably that the "turbulence" is now
necessarily axisymmetric,
but we feel that the results of this section combined with those of Sect.~\ref{res}
can contribute to the case for the more general relevance of the mechanism.

Our standard disc semi-thickness $h=\lambda R$ has $\lambda=0.04$ (so
with $R=10$ kpc, $h=400$ pc), and the computational grid
covers ${\rm r_{\rm min}}\le r \le 1.0$, ${\rm -z_{\rm max}} \le z \le {\rm z_{\rm max}}$, with
$r_{\rm min}=0.05, {\rm z_{\rm max}}=0.3333$ in units of the galactic radius. 
The diffusivity increases asymptotically by a factor of 30 with $|z|$,
and we study a model with no alpha-effect, 
i.e. $R_\alpha=0$, and take $R_\omega=10$.
The random field injections are described by the same parameters as before:
$N_{\rm spot}, r_{\rm spot}, dt_{\rm inj}$, and the interval between
rerandomizations of the spot locations, $T_{\rm spot}=0.3$.
First we display in Fig.~\ref{r,z}a the statistically steady field without
any bulk velocity present. A maybe somewhat surprising feature is the 
apparent presence
of semi-ordered "vortices" in the instantaneous mature fields. We attribute
this feature to the intervals between relocation of the hotspots
being much longer than the field injection intervals (i.e. $T_{\rm spot}>> dt_{\rm inj}$),
so there is sufficient time for diffusion at the boundaries between these
sites to allow the instaneously stronger hotspots to dominate. The height of
the disc may also play a role -- it is known that in spherical thin shell
$\alpha\Omega$ mean field dynamos that there is a tendency for cells to form
with aspect ratio $O(1)$.
In support of this hypothesis we note two subsidiary experiments.\\
1. With just an initial injection of field and no subsequent additions, the field of
course decays. However after some time the  (much weakened) 
field structure reduces to that
of just 3 or 4 vortices.\\
2. If the relocation interval $T_{\rm spot}$ is made slightly shorter than 
the injection interval $dt_{\rm inj}$, the effect almost vanishes.

We now introduce a bulk velocity directed out of the disc. For demonstration
purposes (and simplicity) we take a velocity purely  in the $z$-direction,
$u_z=R_u\tanh(z/h(r))$ (where of course $h(r)=$ const here).
For simplicity, as the model does not involve any mixing of poloidal
and toroidal magnetic fields, below we discuss only the poloidal field 
component.
We show in Figs.~\ref{r,z}b, c, d the steady state  field with $R_u=1, 5, 10$ respectively.
The main features, i.e. the presence of ordered field above the disc,  
appear after a relatively short time, certainly by $\tau<2$.
(We note that the results are substantially similar with $R_u>0$ and $T_{\rm spot} \la dt_{\rm inj}$. 

We also experimented with changing the disc height, and with changing the number of 
hotspots. There was little or no qualitative change in the results.
As might be expected, the magnetic field patterns in the halo region near to the
disc become more pronounced as $R_u$ increases, i.e. with increasing wind velocity. Effects are still very small for $R_u=1$, but the scale of the field
exterior to the disc
reaches a third of the disc radius with $R_u=10$. We note that marked
large-scale structures are produced for significantly smaller values of $R_u$ in 
this case than in the no-$z$ problem of Sect.~\ref{res}, 
but it should be remembered that
the displacement of the structures from the disc is substantially larger
in the no-$z$ case. Again, these figures represent just instantaneous states of
a fluctuating field distribution.

\begin{figure*}
\begin{center}
a \includegraphics[width=0.41\textwidth]{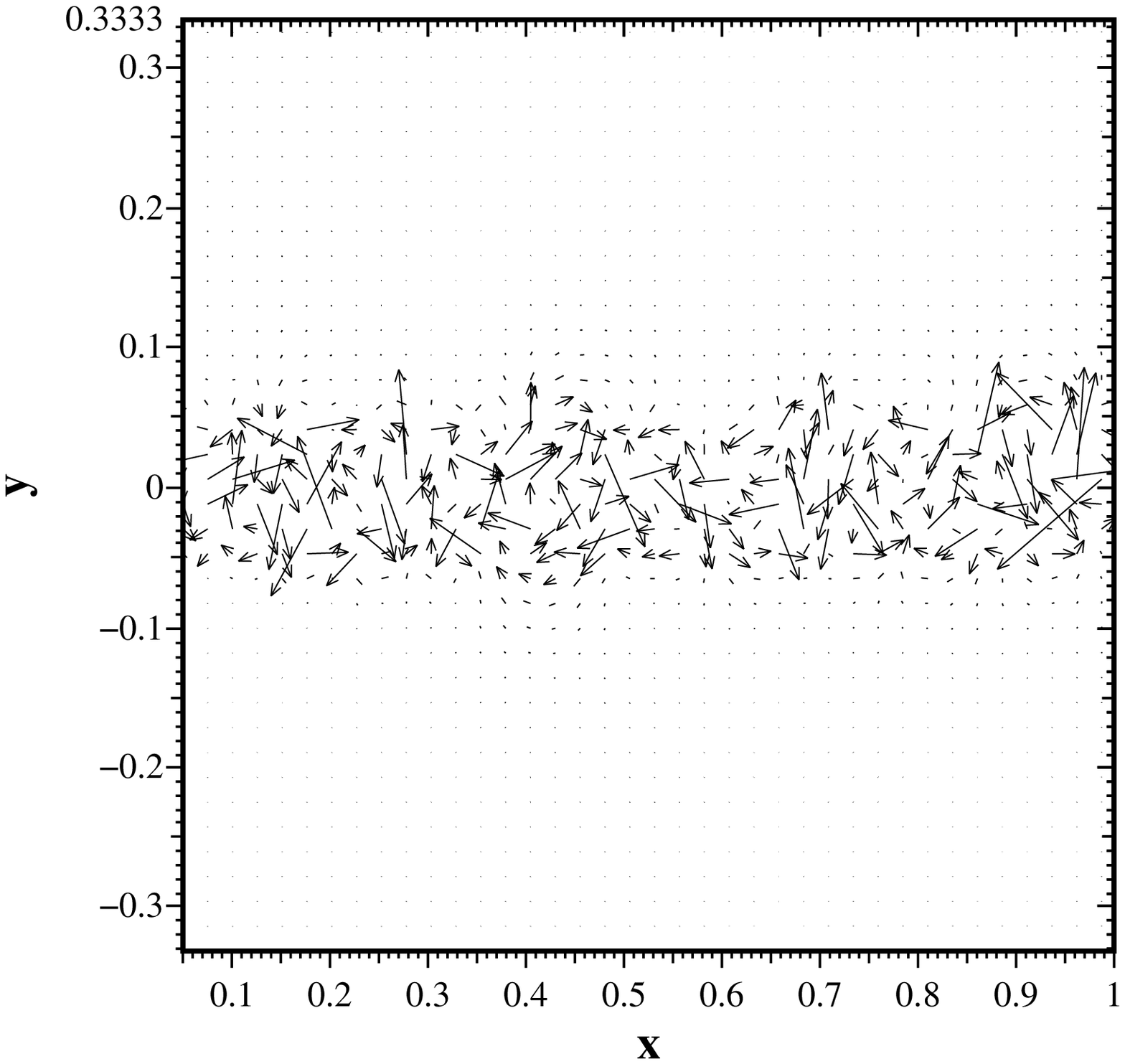} 
b \includegraphics[width=0.41\textwidth]{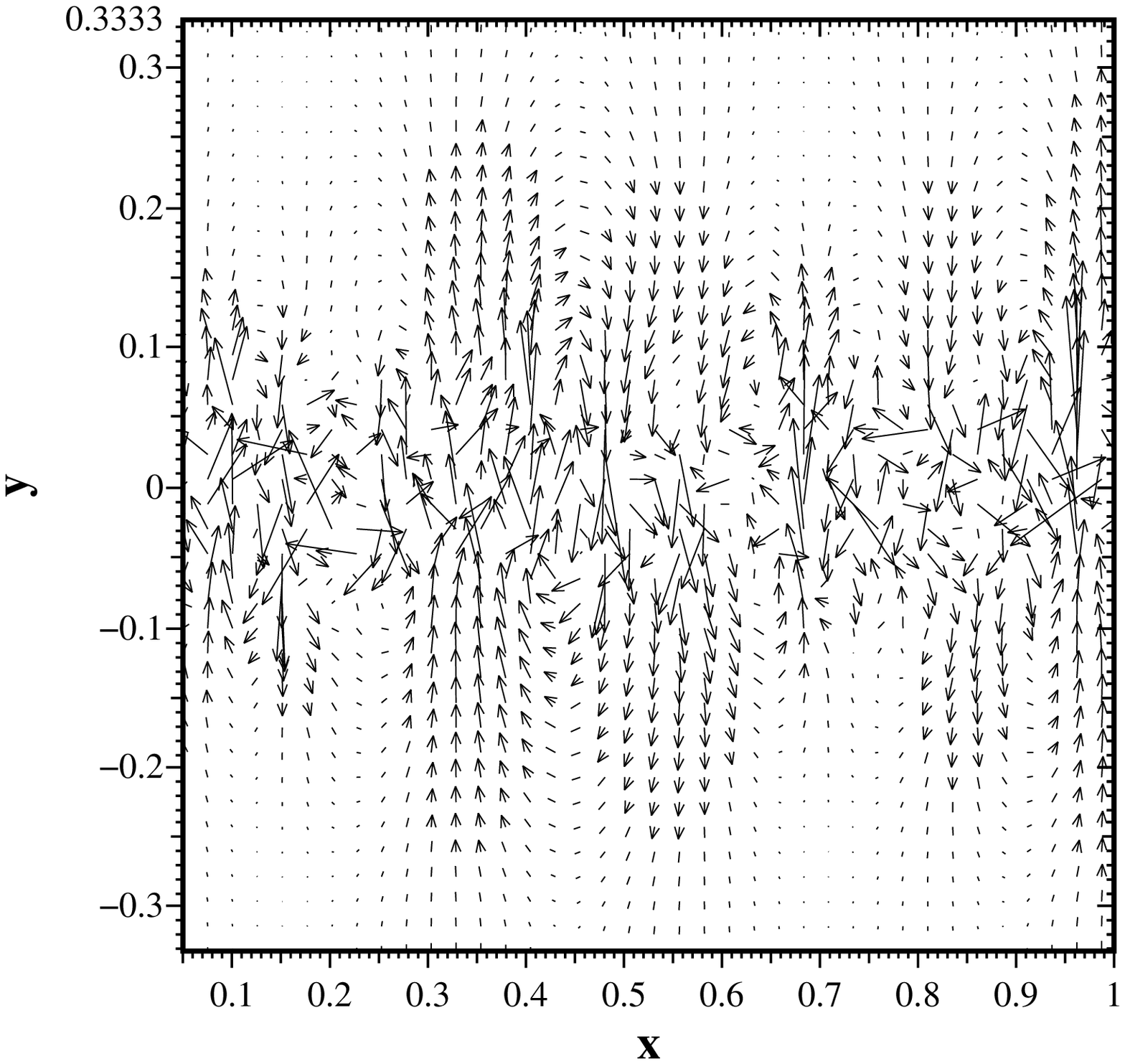} 
\end{center}
\caption[]{ Stationary magnetic field distributions for 
the axisymmetric model in $r,z$ coordinates with $N_{\rm spot}=400, r_{\rm spot}=0.005$ and a) $R_u=0$, b) $R_u=25$.}

\label{nspot400}
\end{figure*}

To avoid the possibility that the out-of-disc field is just an artifact
of the particular implementation of the introduction of the random field, we 
also experimented with random injection parameters that give smaller scale 
disorder,
$r_{\rm spot}=0.005, N_{\rm spot}=400$. 
The statistically steady field in the absence of a bulk velocity, i.e. with $R_u=0$,
is shown in Fig.~\ref{nspot400}a. In panel b of this figure we show the 
statistically steady field when $R_u=25$. It can be seen that results are
quantitatively similar to those obtained previously, albeit using a
somewhat larger (but still plausible) value of $R_u$.

\begin{figure}
\begin{center}
\includegraphics[width=0.41\textwidth]{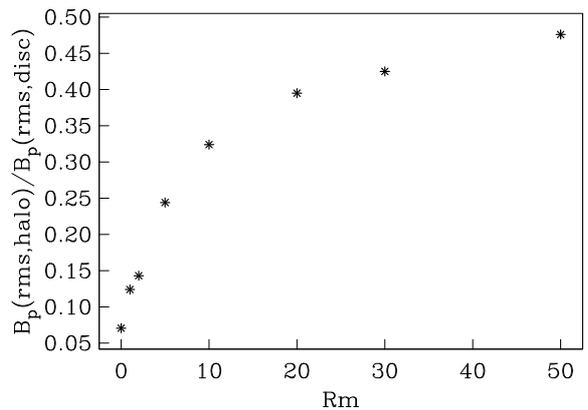} 
\end{center}
\caption[]{ Ratio of r.m.s. poloidal field  in $0.08\le z \le z_{\rm max}$ to that in the disc as a function of $R_u$ for the models with
 $N_{\rm spot}=100$.
}
\label{ratio}
\end{figure}

For the models with $N_{\rm spot}=100$ discussed above, we also calculated 
the ratio of the r.m.s. poloidal field in the halo ($0.08\le z \le z_{\rm max}$)
to that in the disc. These values are plotted in Fig.~\ref{ratio}. Unsurprisingly
the ratio increases with the value of $R_u$. An important effect appears to be amplification
of the halo field  by stretching by the $z$-dependent wind.
The small non-zero value when $R_u=0$ 
arises purely from the large diffusion coefficient above the disc.
(Any small fluctuations are manifestations of the
stochastic nature of the model.)

We feel that these results,  taken with those of Sect.~\ref{res}, provide support for the operation of 
the effect 
discussed independently of the particular geometry in which the model is embedded.

%______________________________________________________________________
\begin{figure*}
\begin{center}
a \includegraphics[width=0.41\textwidth]{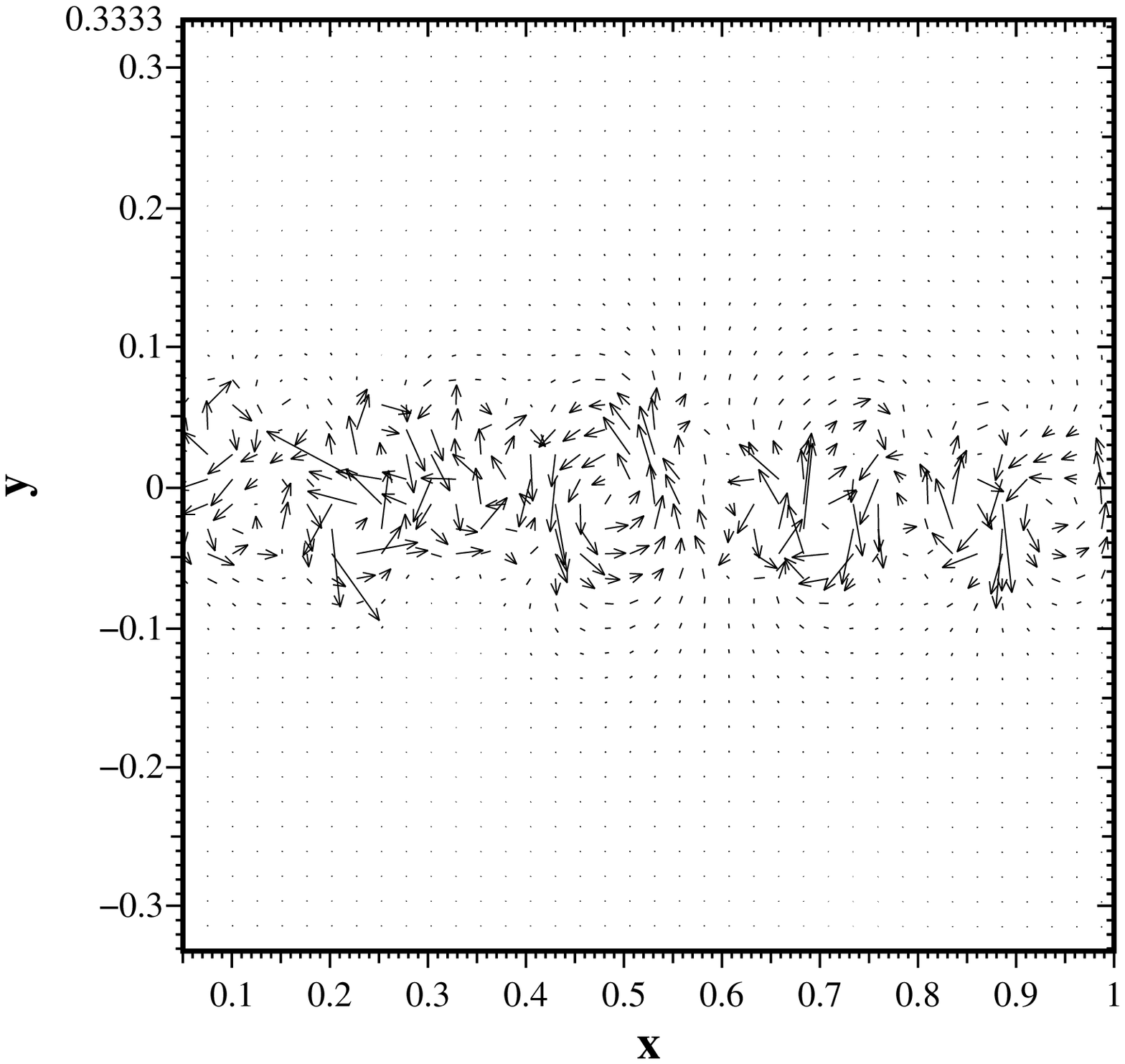}
b \includegraphics[width=0.41\textwidth]{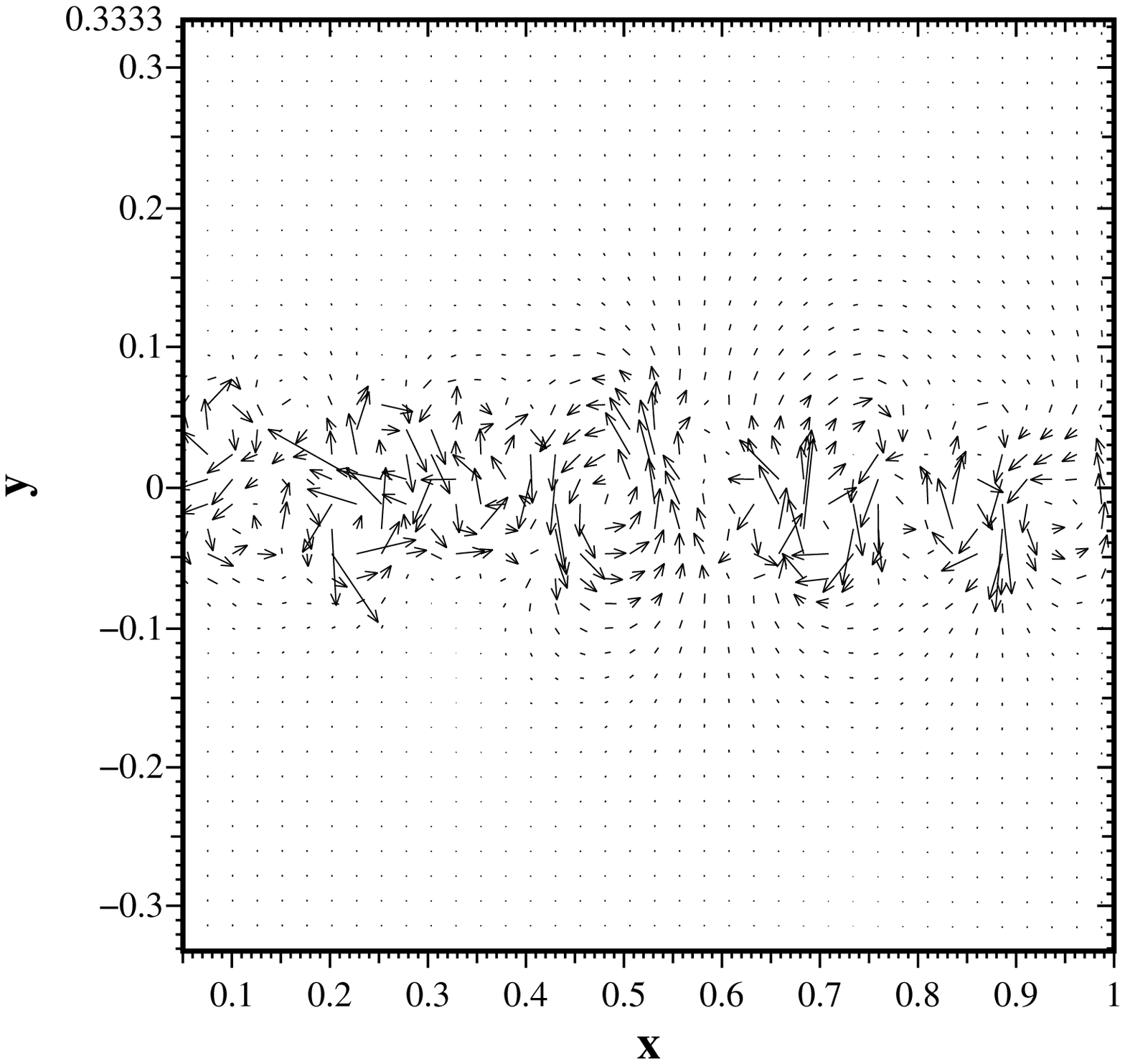}\\
c \includegraphics[width=0.41\textwidth]{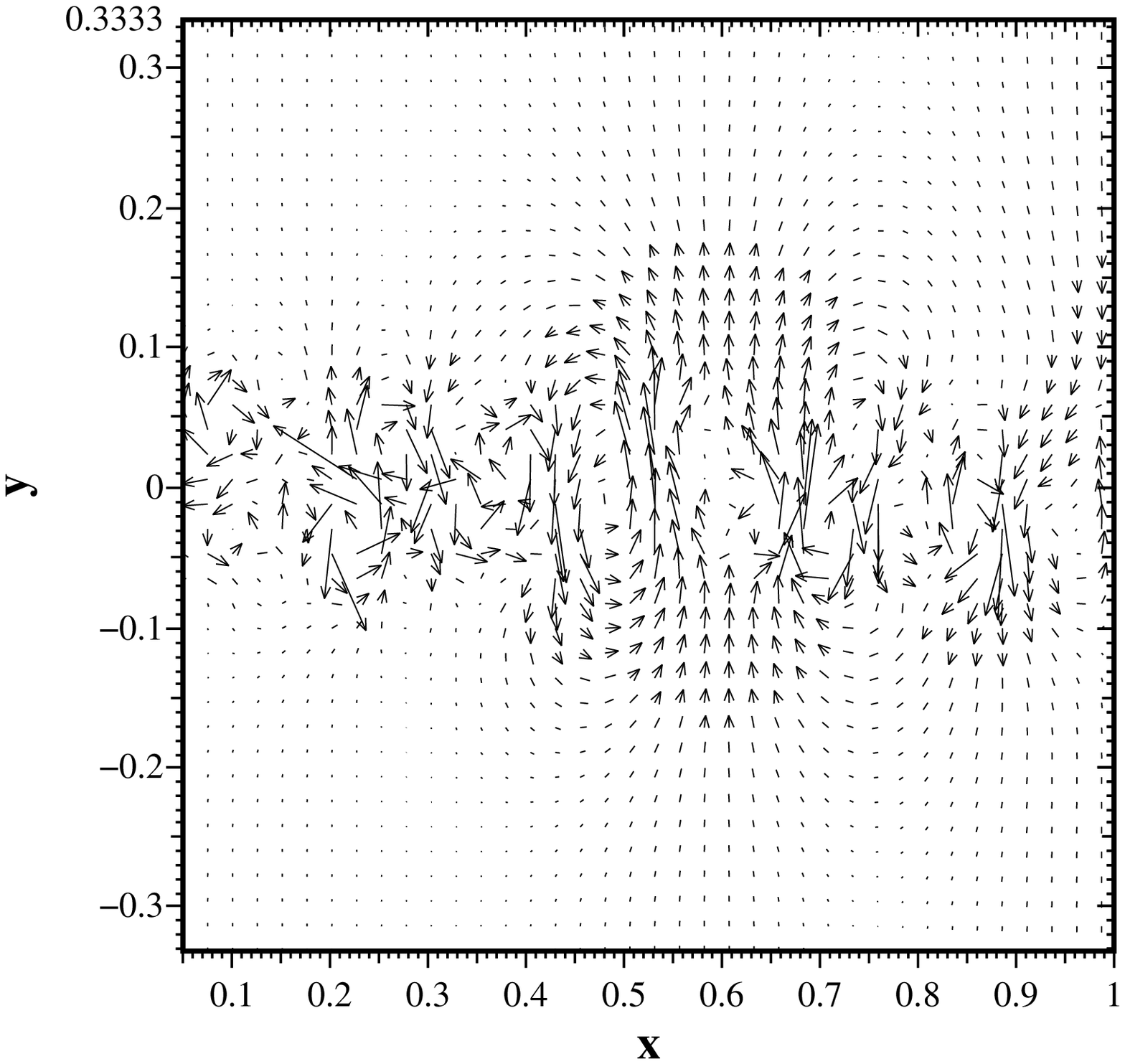}
d \includegraphics[width=0.41\textwidth]{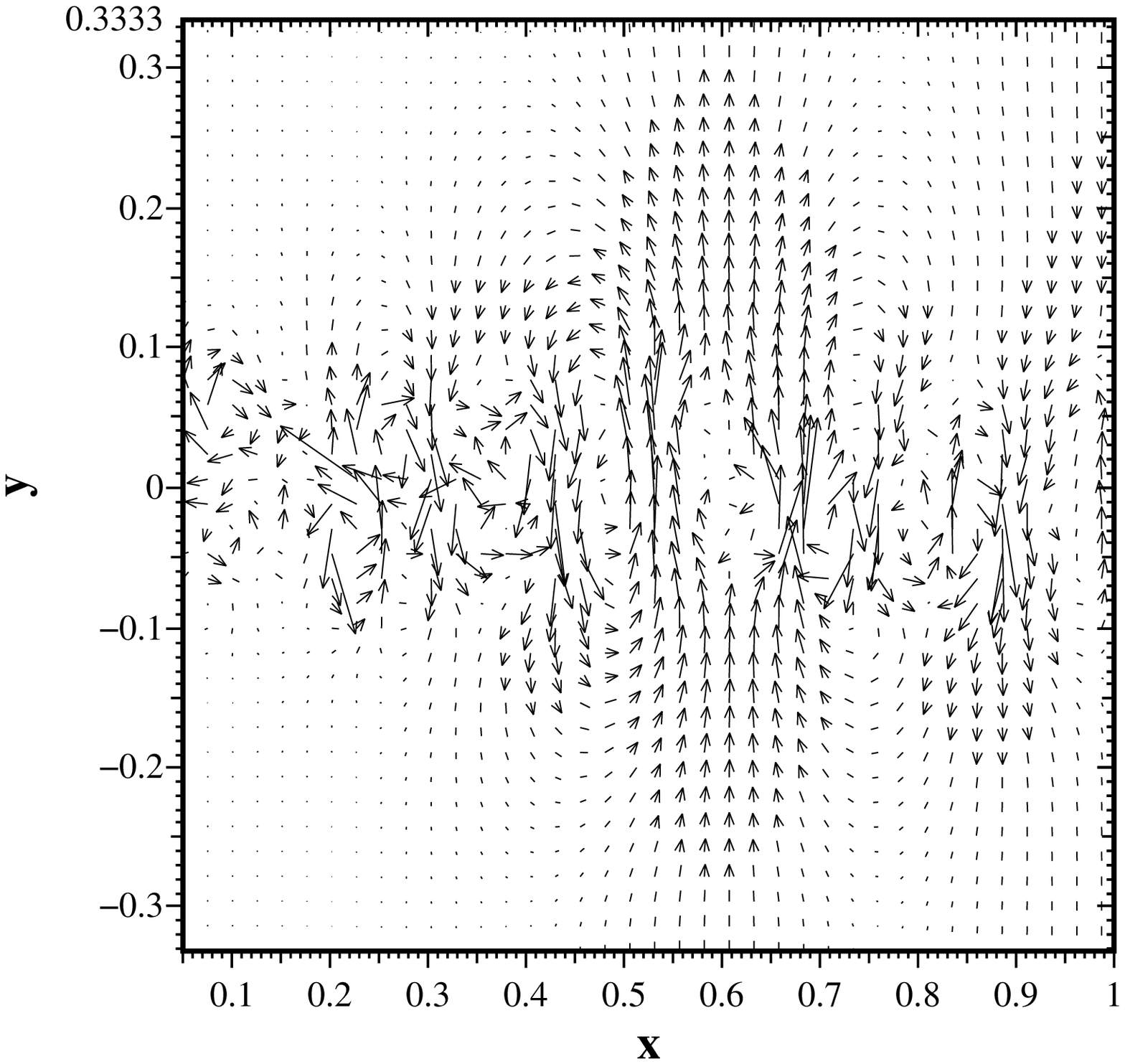}
\end{center}
\caption[]{Stationary magnetic field distributions for 
the axisymmetric model in $r,z$ coordinates for $R_u= 0, 1, 5, 10$ respectively.
The extent of the magnetic field in the region surrounding the disc increases with $R_u$.}

\label{r,z}
\end{figure*}
%______________________________________________________________________
%______________________________________________________________________

\section{Discussion and conclusions} 
\label{disc}

We have presented several numerical models of large-scale magnetic patterns generated by wind transport of 
randomly 
distributed  small-scale magnetic field present in the galactic disc. Even if the large-scale dynamo in the disc remains
subcritical (or equivalently is absent),
the overall magnetic field distribution rapidly reaches a statistically
steady state which resembles a large-scale field 
distribution in the regions external to the disc (the "galactic halo"). 
Amplification caused by stretching by a spatially dependent wind may also play a role.
We believe that this scenario could be relevant 
to producing large-scale magnetic fields in dwarf galaxies, and 
potentially explain certain observations of these objects (Chy\.zy et al. 2016).
Note that Drzazga et al. (2016), 
when discussing observations of a dwarf galaxy NGC~2976 performed in order to 
isolate a large-scale magnetic pattern, discuss the necessity of a mechanism similar to that one suggested here.

Possibly, this scenario (combined with conventional large-scale dynamo action) 
could also be useful for explaining  some
magnetic field structures in stellar atmospheres.

We stress that strictly speaking the magnetic field generated in our modelling in the outermost parts of dwarf 
galaxies 
will still have a vanishing mean. However this fact is mainly of mathematical interest because we only see 
snapshots of 
magnetic field distributions. The magnetic field mean will vanish 
after averaging over a substantial time 
interval, but  the length of this interval is huge compared to the lifetime 
of human beings and substantial even compared to  
galactic timescales.  

We note that in the thin disc model of Sect.~\ref{res} the large-scale pattern is displaced from the disc as far as 
distances comparable with the disc radius. 
The  displacement is large because the wind modelled is quite fast ($R_u=120$ corresponds to $u_0 = 160$ 
km/sec if $\Omega_0 h = 10$ km/sec). Plausibly the actual wind velocity is several times smaller  and so, correspondingly, would be
the displacement. 
The point however is that if we consider displacement of the 
large-scale field pattern in the direction orthogonal to the disc 
then a displacement by 1/5 of the galactic radius 
seems sufficient, as galactic discs are thin compared to their radii. 
This is supported by the models of Sect.~\ref{axi}.
In Sect.~\ref{res} we enhanced the effect to make it pronounced while retaining
the 2D format of the problem.

Presumably, the effect observed depends on the number of injections being 
quite, but not very, large.
In this sense the effect discussed looks similar to the proposed 
explanation for the presence of a nonvanishing magnetic 
moment during solar magnetic field inversions (Moss et al. 2013). 
This magnetic moment randomly 
migrates from being almost parallel to the rotation axis to being almost 
antiparallel (Pipin et al. 2014). The 
net magnetic moment appears due to the finite number of active regions.  We agree that if the number of 
injection sites were much larger and, certainly if anywhere near to comparable with, say, Avogadro's number, the resulting large-scale pattern would be averaged
out. 

We also note that the mechanism suggested has some features in common with 
that suggested by Chakrabarti et al. (1994) which, however, 
is addressed to a different astrophysical situation  (see also Cho 2014). 
A related mechanism, in which turbulent diffusion plays an important role,
 was discussed in a galactic context by Blackman (1998).

Our simulations are intended only as a demonstration of a physical effect, and 
are not specifically directed to modelling any specific object or particular 
type of galaxy.  Thus we have not attempted to model observational consequences of the models, such as polarization patterns. 
Such an undertaking would require much more computationally
demanding 3D modelling and corresponding adjustment of physical parameters, and would need  to be addressed separately. 
We note however that our results appear broadly compatible with what is known about magnetic fields in dwarf galaxies.
We are content here to demonstrate what we feel may be an interesting physical process.

\begin{acknowledgements}
DS is grateful to RFBR for financial support under grant 15-02-01407. 
Useful discussions with Olga Silchenko are acknowledged.
Comments from an anonymous referee were valuable.

\end{acknowledgements} 

%______________________________________________________________________
%______________________________________________________________________

\end{document}